\documentclass{article}
\usepackage{spconf,amsmath,graphicx}
\usepackage{graphics}
\usepackage{algorithm}
\usepackage{algpseudocode}
\usepackage{siunitx}
\usepackage[subtle]{savetrees}
\usepackage{xcolor, soul}

\title{Particle Flow Gaussian Sum Particle Filter }
\name{Karthik Comandur$^{\dagger}$, Yunpeng Li$^{\star}$ and Santosh Nannuru $^{\dagger}$}

\address{$^{\dagger}$ Signal Processing and Communication Research Centre, IIIT Hyderabad, India. \\
$^{\star}$Department of Computer Science, University of Surrey, UK.}

\begin{document}
%
\maketitle
\begin{abstract}
The particle flow Gaussian particle filter (PFGPF) uses an invertible particle flow to generate a proposal density. It approximates the predictive and posterior distributions as Gaussian densities. In this paper, we use a bank of PFGPF filters to construct a Particle flow Gaussian sum particle filter (PFGSPF), which approximates the prediction and posterior as Gaussian mixture model. This approximation is useful in complex estimation problems where a single Gaussian approximation is inadequate. We compare the performance of this proposed filter with the PFGPF and others in challenging numerical simulations.
\end{abstract}
\section{ Introduction}
\label{sec:intro}

Sequential Bayesian state estimation is performed by tracking the posterior distribution. Particle filters 
\cite{gordon1993novel} 
approximate the posterior distribution using a set of particles and their associated weights. In high-dimensional non-linear state space models, particle filters suffer from weight degeneracy \cite{bickel2008sharp,bengtsson2008curse,snyder2008obstacles}, which results in poor approximation of the posterior. Different particle filters have been proposed to overcome the weight degeneracy issue \cite{arulampalam2002tutorial,pitt1999filtering,doucet2000s}.
Though these filters are effective in many settings, their applicability is often restricted due to their algorithmic assumptions.



One effective direction to tackle weight degeneracy is to perform Markov Chain Monte Carlo (MCMC) sampling after the resampling step to improve the diversity of the particles~\cite{berzuini1997dynamic,gilks2001following,godsill2001improvement,musso2001improving,brockwell2010sequentially,septier2015langevin}. A large number of computationally expensive MCMC iterations are often required to overcome weight degeneracy.
An alternative approach is through a particle flow procedure. Particle flow filters \cite{daum2007nonlinear,daum2008particle,daum2009gradient,daum2010exact,daum2013particle,daum2014seven}
migrate particles from the prior to the posterior following a flow equation. There is no weight degeneracy issue as no resampling is performed. These filters involve various model assumptions and approximations in their numerical implementations, which leads to statistical inconsistency.


To inherit the desirable statistical properties of particle filters while incorporating particle flow, invertible particle flow procedures are constructed to generate a proposal distribution within the particle filter or sequential MCMC framework~\cite{li2017particle,li2019}. The invertible particle flows proposed in~\cite{li2017particle} were adapted from the exact Daum and Huang (EDH) filter \cite{daum2010exact} and the localised exact Daum and Huang (LEDH) filter \cite{ding2012implementation} respectively. The numerical implementations of the invertible particle flow suggest that the resulted proposal distribution can be Gaussian or local Gaussian approximation of the posterior~\cite{li2019}.


Another type of filters that circumvent the resampling step is the Gaussian particle filter (GPF) \cite{article}
which approximates both predictive and posterior distributions as Gaussian densities. A new set of particles are drawn from the Gaussian approximation of the posterior before each iteration. Recently, an inverted particle flow is incorporated into a Gaussian particle filter in the particle flow Gaussian particle filter (PFGPF) \cite{PFGPF2022} to construct a modified proposal density. Similar to the GPF, there is no resampling in the PFGPF, which avoids weight degeneracy.

To better tackle state space models with multi-modal distributions, Gaussian mixture models have been adopted to derive variants of the Gaussian particle filter or particle flow-based filters.
The Gaussian Sum Particle filter (GSPF) \cite{GSPF2001} is built using banks of the GPF. The predictive and posterior distributions are approximated as Gaussian mixture distributions. 
Particle flows with various Gaussian mixture assumptions were derived in \cite{khan2016log,pal2017,pal2018,li2019} for incorporation into the particle filtering or sequential MCMC frameworks. However, despite that these particle flow procedures in effect generate mixture of Gaussian distributions, none of them have been incorporated into the Gaussian sum particle filtering framework. 

In this paper, we propose the particle flow Gaussian Sum particle flow filter (PFGSPF), which is constructed using banks of PFGPF. The predictive and posterior distributions are approximated as Gaussian mixture densities, as Gaussian mixture can be a better approximation of the posterior than a single Gaussian in complex, high-dimensional filtering scenarios.
The main contributions of this paper include:
(i) we incorporate the effectiveness of invertible particle flow into an encompassing Gaussian sum particle filtering framework to address weight degeneracy;
(ii) we introduce an \textit{effective number of Gaussians} ($G_\text{eff}$) to find the number of Gaussian densities significant in the Gaussian mixture distribution;
(iii) we show superior empirical performance of the PFGSPF in two challenging numerical simulations.

The structure of this paper is as follows: 
Section \ref{sec:Problem Statement} introduces the problem statement. In Section \ref{sec:GSPF}, we review Gaussian sum particle filtering. We propose particle flow Gaussian sum particle filter in Section \ref{sec:PFGSPF} and present simulation results in Section \ref{sec:SimandResults}. The paper is concluded in Section \ref{sec:conclusion}.

\section{Problem Statement}
\label{sec:Problem Statement}


The unobserved state $x_t$ at time step $t$ of the non-linear system is estimated by tracking the posterior density $p(x_t|z_{1:t})$ over time where $z_{1:t} = \{z_1,...,z_t\}$ is the set of observations collected up to $t$. The dynamic model and the observation model are given by
\begin{align} 
x_0 &\sim {p}_0(x)\,\,,\\
x_t &= g_t(x_{t-1},v_t)\,\,, \; \quad {t = 1,2,\ldots}  \label{eq:dynamic_model} \,, \\ 
z_t &= h_t(x_t,w_t)\,\,, \; \quad {t = 1,2,\ldots}  \label{eq:measurement_model} \,,
\end{align}
where $p_0(x)$ is the initial state distribution, $g_t(\cdot)$ is the state transition function, and $h_t(\cdot)$ is the observation model which generates the observations $z_t$. The process noise and observation noise are $v_t$ and $w_t$, respectively. We assume that $g_t(\cdot,0)$ is bounded and $h_t(\cdot,0)$ is a $C^1$ function, i.e. $h_t(\cdot,0)$ is differentiable everywhere and its derivatives are continuous.

\section{Related work}
\label{sec:GSPF}

Gaussian sum particle filters~\cite{GSPF2001} approximate the predictive and posterior distributions as a Gaussian mixture. At time step $t-1$ the posterior is approximated by $p(x_{t-1}|z_{1:t-1}) \approx \sum_{j=1}^{G} \alpha^j_{t-1}{\mathcal{N}(x_{t-1};\mu_{t-1}^j,\Sigma_{t-1}^j)}$ where $G$ is the total number of Gaussian distributions in the mixture. 
For each Gaussian in the mixture, the GPF is applied. The prediction and update steps for the Gaussian components are discussed in the following subsections. Please refer to \cite{GSPF2001} for details.

\textbf{The prediction step}: A set of particles $\{{x}^{ij}_{t-1}\}^{N^{*}_p}_{i=1}$ are drawn from the $j^\text{th}$ Gaussian distribution $\mathcal{N}(x^{ij}_{t-1};\mu^j_{t-1},\Sigma^j_{t-1})$. This particle set is propagated through the dynamic model $g_t(\cdot)$ to generate the predictive particles, ${x}^{ij}_{t} = {g}_t({x}^{ij}_{t-1},v_t)$. The empirical mean and covariance of this set of particles are used for Gaussian approximation, $\mathcal{N}(\cdot\,;\overline{\mu}^{j}_{t},\overline{\Sigma}^{j}_{t})$, of the predictive distribution:
\begin{align} 
{\overline{\mu}}^{j}_{t} &= \frac{1}{N^{*}_{p}} {\sum}^{N^{*}_{p}}_{i=1} {x}^{ij}_{t} \,\,, \label{eq:GSPF_prediction_mean} \\ 
{\overline{\Sigma}}^{j}_{t} &= \frac{1}{N^{*}_{p}} {\sum}^{N^{*}_{p}}_{i=1} ({x}^{ij}_{t}-{\overline{\mu}}^{j}_{t}) {({x}^{ij}_{t}-{\overline{\mu}}^{j}_{t})}^{T}\,\,. \label{eq:GSPF_prediction_cov}
\end{align} 
The predictive distribution is approximated as $p(x_{t}|z_{1:t-1}) \approx \sum_{j=1}^{G} \alpha^j_{t-1}{\mathcal{N}(x_{t};\overline{\mu}^{j}_{t},\overline{\Sigma}^{j}_{t})}$.

\textbf{The update step}: The importance weight of the $i^{th}$ particle in the $j^\text{th}$ Gaussian is computed as
\begin{align}
{w}^{ij}_{t} &\propto \frac{\mathcal{N}(x^{ij}_{t};\overline{\mu}^{j}_{t},\overline{\Sigma}^{j}_{t})p(z_t|{x}^{ij}_{t})}{\pi({x}^{ij}_{t}|z_{1:t})}\,\,.
\label{eq:GSPF_weight}
\end{align}
where $\pi(\cdot|z_{1:t})$ is the proposal density.
A Gaussian density ${\mathcal{N}(x_{t};{\mu}^{j}_{t},{\Sigma}^{j}_{t})}$ is approximated from this particle set where, $\mu^{j}_{t}={\sum}^{N^{*}_{p}}_{i=1} \underline{w}^{ij}_{t} {x}^{ij}_{t}$ and $\Sigma^{j}_{t}=
{\sum}^{N^{*}_{p}}_{i=1} \underline{w}^{ij}_{t}({x}^{ij}_{t}-{\mu}^{j}_{t}) {({x}^{ij}_{t}-{\mu}^{j}_{t})}^{T}$ are the weighted mean and covariance of the $j^\text{th}$ Gaussian, respectively. We use $\underline{w}^{ij}_{t}$ to denote normalized ${w}^{ij}_{t}$.

The posterior distribution is approximated as $p(x_{t}|z_{1:t}) \approx \sum_{j=1}^{G} \alpha^j_{t}{\mathcal{N}(x_{t};{\mu}^{j}_{t},{\Sigma}^{j}_{t})}$, where the mixing proportion of the $j^\text{th}$ Gaussian $\alpha^j_{t}$ are updated and normalized as follows,
\begin{align}
\tilde{\alpha}^j_{t} &= \alpha^j_{t-1}\frac{\sum_{i=1}^{N^{*}_p}{w^{ij}_{t}}}{\sum_{j=1}^{G}\sum_{i=1}^{N^{*}_p}{w^{ij}_{t}}} \,, \quad
\alpha^j_{t} = \frac{\tilde{\alpha}^j_{t}}{\sum_{j=1}^{G}\tilde{\alpha}^j_{t}}. 
\label{eq:alpha_update_1}
\end{align}

\section{Particle Flow Gaussian Sum Particle \\ Filter}
\label{sec:PFGSPF}

In the proposed particle flow Gaussian sum particle filter (PFGSPF), the posterior is approximated by a Gaussian sum distribution. At time step $t-1$, the posterior is approximated by $p(x_{t-1}|z_{1:t-1}) \approx \sum_{j=1}^{G} \alpha^j_{t-1}{\mathcal{N}(x_{t-1};\mu_{t-1}^j,\Sigma_{t-1}^j)}$ where $G$ is the total number of Gaussian distributions in the mixture. The PFGSPF applies the particle flow Gaussian particle filter (PFGPF)~\cite{PFGPF2022} for each Gaussian component as follows.

\subsection{The prediction step}
Particles $\{{x}^{ij}_{t-1}\}^{N^{*}_{p}}_{i=1}$ drawn from the $j^{\text{th}}$ Gaussian distribution $\mathcal{N}(x^{ij}_{t-1};\mu^j_{t-1},\Sigma^j_{t-1})$ are propagated through the dynamic model $g_t(\cdot)$ to generate the predictive particles,
\begin{align}
{\eta}^{ij}_{0} &= {g}_t({x}^{ij}_{t-1},v_t).
\end{align}
The empirical mean and empirical covariance of this set of particles $\{\eta^{ij}_{0}\}_{i=1}^{N^{*}_{p}}$ are computed using \eqref{eq:GSPF_prediction_mean} and \eqref{eq:GSPF_prediction_cov} by replacing $x^{ij}_{t}$ with $\eta^{ij}_{0}$. These parameters are used to construct the Gaussian approximation, $\mathcal{N}(\cdot\,;\overline{\mu}^{j}_{t},\overline{\Sigma}^{j}_{t})$, of the predictive distribution.


For each Gaussian, the invertible particle flow (LEDH)~\cite{ding2012implementation, li2017particle} is applied to the particle set $\{\eta^{ij}_{0}\}_{i=1}^{N^{*}_{p}}$ to generate the migrated particles $\{\eta^{ij}_{1}\}_{i=1}^{N^{*}_{p}}$ to form the proposal density.
The functional mapping $T(\cdot)$ for $i$-th particle (we ignore the superscript $j$ below) using the LEDH flow is given as
\begin{align}
     \eta^i_{\lambda_l} &= f^{i}_{\lambda_l}(\eta^i_{\lambda_{l-1}}), \quad  l=1,2,...N_\lambda  \\
    &= \eta^i_{\lambda_{l-1}} + \epsilon_{l}(A^{i}(\lambda_l) \eta^i_{\lambda_{l-1}} + b^{i}(\lambda_l)),
    \label{eq: LEDH_map}
\end{align}
where $\lambda \in [0,1]$ is the pseudo-time interval, $[\lambda_1, \lambda_2,.....,\lambda_{N_\lambda}]$ are the $N_\lambda$ pseudo time steps, $\epsilon_l = \lambda_{l}- \lambda_{l-1}$ is the step size and $\sum^{N_\lambda}_{l=1} \epsilon_{l} = 1$. The flow parameters $A^{i}(\lambda)$ and $b^{i} (\lambda)$ for the $i$-th particle are computed as follows:
\begin{align}
&{A}^i(\lambda) = -\frac{1}{2}P{{H}^i(\lambda)}^T{(\lambda {H}^i(\lambda)P{{H}^i(\lambda)}^T+R)}^{-1} {H}^i(\lambda) \,\,, \label{eq:ledh_eqA} \\
&{b}^i(\lambda) = (I+2\lambda {A}^i(\lambda)) \times \nonumber\\
& \; [(I+\lambda {A}^i(\lambda)) P{{H}^i(\lambda)}^T{R}^{-1}(z-{e}^i(\lambda) + {A}^i(\lambda)\overline{\eta_0}]\,\,.
\label{eq:ledh_eqB}
\end{align}
Here $\overline{\eta_{0}}$ is the predicted mean, $P$ is the predicted covariance, $R$ is the observation covariance and $z$ is the observation. The linearization of the observation model is performed at each individual particle as 
${H}^i(\lambda)=\frac{\partial h(\eta,0)}{\partial \eta} {\Big|}_{\eta={{\eta^i_\lambda}}}$ and $e^{i}(\lambda) = h(\eta^{i}_{\lambda},0)-H^{i}(\lambda) \eta^{i}_{\lambda}$.


\subsection{The update step}
Following the PFPF \cite{li2017particle} and PFGPF \cite{PFGPF2022}, the proposal density $\pi(\cdot)$ is computed as
\begin{align}
\pi({\eta}^{ij}_{1}|z_{0:t}) &= \frac{\mathcal{N}(\eta^{ij}_0;\overline{\mu}^j_{t},\overline{\Sigma}^j_{t})}{|\dot{T}{({\eta}^{ij}_{0};z_{t},{x}^{i}_{t-1})}|}\,\,,
\label{eq:PFGPF_IS}
\end{align}
where $|\dot{T}(\cdot)|$ is the determinant of the Jacobian of the mapping function ${T(\cdot)}$ between the predicted particles and migrated particles. The importance weight of the $i^{\text{th}}$ particle ${\eta}^{ij}_{1}$ is computed as
%
%
\begin{align}
{w}^{ij}_{t} &\propto \frac{\mathcal{N}(\eta^{ij}_{1};\overline{\mu}^{j}_{t},\overline{\Sigma}^{j}_{t})p(z_t|{\eta}^{ij}_{1})|\dot{T}{({\eta}^{ij}_{0};z_{t},{x}^{ij}_{t-1})}|}{\mathcal{N}(\eta^{ij}_{0};\overline{\mu}^{j}_{t},\overline{\Sigma}^{j}_{t})}\,\,.
\label{eq:PFGPF_weight}
\end{align}
As in the PFGPF, the Gaussian distribution $\mathcal{N}(\cdot ;\mu^{j}_{t},\Sigma^{j}_{t})$ is used to approximate the posterior distribution after the weight update step. Here, $\mu^{j}_{t}={\sum}^{N^{*}_{p}}_{i=1} \underline{{w}}^{ij}_{t} {\eta}^{ij}_{1}$ and $\Sigma^{j}_{t}=
{\sum}^{N^{*}_{p}}_{i=1} \underline{{w}}^{ij}_{t}({\eta}^{ij}_{1}-{\mu}^{j}_{t}) {({\eta}^{ij}_{1}-{\mu}^{j}_{t})}^{T}$, where $\underline{w}^{ij}_{t}$ are the normalized ${w}^{ij}_{t}$.


All the $G$ Gaussians are now combined to obtain the posterior. The mixing proportions are updated and normalized following \eqref{eq:alpha_update_1}. The Gaussian sum distribution $\sum_{j=1}^{G} \alpha^j_{t}{\mathcal{N}(x_{t};\mu_{t}^j,\Sigma_{t}^j)}$ approximates the posterior. The estimated state $\hat{x}_{t}$ is computed as follows,
\begin{align}
\hat{x}_{t} = \sum_{j=1}^{G}\alpha^j_{t}\mu_{t}^j \,.
\end{align}
The PFGSPF uses a total of $N_p = N^*_p \times G$ particles in its implementation as there are $N_p^*$ particles per Gaussian and a total of $G$ Gaussians in the mixture. Algorithm \ref{algorithm_1} outlines the proposed filter where we use a bank of PFGPF to construct the PFGSPF filter.

\begin{algorithm}
\linespread{1.1}\selectfont
\caption{{ Particle flow Gaussian sum particle filter}}
\label{algorithm_1}
 {\bf Input:} The posterior at time $t-1$ \newline
$p(x_{t-1}|z_{1:t-1}) = \sum_{j=1}^{G}\alpha^j_{t-1}{\mathcal{N}(x_{t-1};\mu_{t-1}^j,\Sigma_{t-1}^j)}$
\vspace{0.1cm}
\hrule
\vspace{0.1cm}
\begin{algorithmic}[1]
\label{alg:PFGSPF}
\For {$j = 1$ to $G$}
\State Apply PFGPF to ${\mathcal{N}(x_{t-1};\mu_{t-1}^j,\Sigma_{t-1}^j)}$;
\State ${\mathcal{N}(x_{t-1};\mu_{t-1}^j,\Sigma_{t-1}^j)} \xrightarrow[]{\text{PFGPF}}  {\mathcal{N}(x_{t};\mu_{t}^j,\Sigma_{t}^j)}$;
\EndFor
\For {$j = 1$ to $G$}
\State Update the mixing proportions using \eqref{eq:alpha_update_1};
\EndFor
\State Normalize the mixing proportions using \eqref{eq:alpha_update_1};
\end{algorithmic}
\vspace{0.1cm}
\hrule
\vspace{0.1cm}
{\bf Outputs:} Posterior: $p(x_{t}|z_{1:t}) \approx \sum_{j=1}^{G} \alpha^j_{t}{\mathcal{N}(x_{t};\mu_{t}^j,\Sigma_{t}^j)}$.
State estimation: $\hat{x}_{t} = \sum_{j=1}^{G}\alpha^j_{t}\mu_{t}^j$
\end{algorithm}

\subsection{Effective number of Gaussians}
Though the number of Gaussians is fixed ($G$) over an entire simulation, their mixing proportions change over time, reflecting the complexity of the distribution as time progresses. Some of the mixing proportions may have very low value indicating that the particular Gaussian has negligible contribution to the overall posterior density. As a measure of complexity of the distribution, we propose the \textit{effective number} of Gaussians $G_{\text{eff},t}$ at time step $t$ defined as follows, 
\begin{align}
G_{\text{eff},t} = \frac{1}{\sum_{j=1}^{G}(\alpha^j_{t})^2}\,\,.
\end{align}
It is trivial to show that when only one Gaussian is signficiant (with $\alpha \approx 1$), the $G_{\text{eff}}$ is close to 1 and when all the $G$ Gaussians have equal weights, the $G_{\text{eff}}$ is equal to $G$.

\section{Simulations and Results}
\label{sec:SimandResults}

We evaluate the proposed algorithm in two challenging numerical simulations -- multi-target tracking using acoustic sensors and a large spatial sensor network model. The performance of the proposed PFGSPF is compared with a number of related algorithms including the EDH~\cite{daum2010exact}, the LEDH~\cite{ding2012implementation}, the PFPF (EDH) and the PFPF (LEDH)~\cite{li2017particle}, and the PFGPF~\cite{PFGPF2022}.

\subsection{Multi-target acoustic tracking}

The simulated acoustic tracking setup of \cite{li2017particle,hlinka2011distributed} involves $M=4$ acoustic targets with state dynamics $x^m_{t} = F x^m_{t-1} + v^m_{t}$, where the state of the $m^{\text{th}}$ target $x^m_{t} = [{\rm x}^m_{t},{\rm y}^m_{t},\dot{{\rm x}}^m_{t},\dot{{\rm y}}^m_{t}]$ consists of the position and velocity components. The process noise $v^m_t \sim \mathcal{N}(0,V)$.
The $s^{\text{th}}$ sensor, located at position $r^{s}$, records the measurement
\begin{align}
\overline{z}^s(x_t) = \sum^{M}_{m=1} \frac{\psi}{{\| {[{\rm x}^m_t,{\rm y}^m_t]}^T - r^s \|}_2 + d_0} \,\,, \label{eq_sensor_amp}  
\end{align}
where $\|\cdot\|_2$ is the Euclidean norm, $\psi$ is the amplitude of the sound emitted by the targets, and $d_0 = 0.1$. There are $N_{s} = 25$ sensors located in the given region. The measurement noise follows $\mathcal{N}(0, \sigma^2_w)$ with variance $\sigma^2_w = 0.01$. We simulate 100 trajectories with a constant velocity model and the corresponding measurements. Each algorithm runs $5$ times with different initial distribution. All other parameter values are the same as those in \cite{PFGPF2022}.

The numerical simulations are carried out for $N_p=500$ and $N_p=2500$ particles.
For the PFGSPF, $N_p = N^{*}_{p} \times G$ and we vary $G$ from $1$ to $5$. We set $N^{*}_{p} = 100$ and $N^{*}_{p} = 500$.
Table \ref{tab:acoustic_without_invalid_tracks} reports the average optimal mass
transfer (OMAT) error and standard deviation of various filters. The error is estimated for tracks excluding the lost tracks (i.e., tracks which have an OMAT error $>$ \SI{2}{\meter}). The number of lost tracks (LT) is also shown in the table. We observe that at $N_p=2500$, the proposed filter PFGSPF outperforms all the other filters with the smallest mean OMAT values and the lowest standard deviation compared to the other filters. We also observe that when $N^{*}_p$ is small, the performance of the PFGSPF is not much affected with increase in number of Gaussians $G$, but as we increase $N^{*}_{p}$ , the performance of the PFGSPF improves as we increase the number of Gaussians $G$.


\begin{table}[t]
\small
\centering
    \begin{tabular}{|c|c|c|c|c|}
    \hline
    {Algorithm} &
       \multicolumn{2}{c|}{$N_p = 500$}&
       \multicolumn{2}{c|}{$N_p = 2500$}
       \\
    \cline{2-5}
    & mean $\pm$ sd & $\#$LT & mean $\pm$ sd & $\#$LT \\
    \hline
    EDH & 0.92$\pm$ 0.39 & 212 & 0.90 $\pm$ 0.37&
    212\\
    \hline
    LEDH & 1.28$\pm$ 0.32 & 111& 1.29$\pm$ 0.33 & 103  \\
    \hline
    PFPF (EDH) & 0.97$\pm$ 0.37 &211 & 0.98$\pm$ 0.39 & 213\\
    \hline
    PFPF (LEDH) & 0.73$\pm$ 0.19 & 5 & 0.74$\pm$ 0.22 & 4 \\
    \hline
    PFGPF & 0.72$\pm$ 0.16 &2& 0.77$\pm$ 0.19 & 0 \\
    \hline
    \hline
    & $N_p$=$N^{*}_{p} \times G$ & & $N_p$=$N^{*}_{p} \times G$ & \\
    \hline
    PFGSPF & ($N^{*}_{p}=100$) & $\#$LT & $(N^{*}_{p}=500$)  & $\#$LT \\
    \hline
    $G=1$ & 0.76 $\pm$ 0.23 & 42 & 0.72 $\pm$ 0.16 &  4\\
    \hline
    $G=2$ & 0.75 $\pm$ 0.25 & 40& 0.70 $\pm$ 0.16 &2 \\
    \hline
    $G=3$ & 0.76 $\pm$ 0.26 & 22& 0.69 $\pm$ 0.13 & 1\\
    \hline
    $G=4$ & 0.76 $\pm$ 0.26 & 33 & 0.69
     $\pm$ 0.16 & 1\\
    \hline
    $G=5$ & 0.75  $\pm$ 0.24 & 29& 0.68  $\pm$ 0.13 & 3\\
    \hline
    
    \end{tabular}
\caption{\label{tab:acoustic_without_invalid_tracks} OMAT errors (m) recorded for the acoustic tracking simulation, excluding lost tracks (LT) defined as the estimated trajectories with average tracking error $>$ \SI{2}{\meter}.}
\end{table}


\subsection{Large spatial sensor networks: Skewed-t dynamic model and count measurements}

\begin{figure}[b]
\centering
\includegraphics[width=0.45\textwidth]{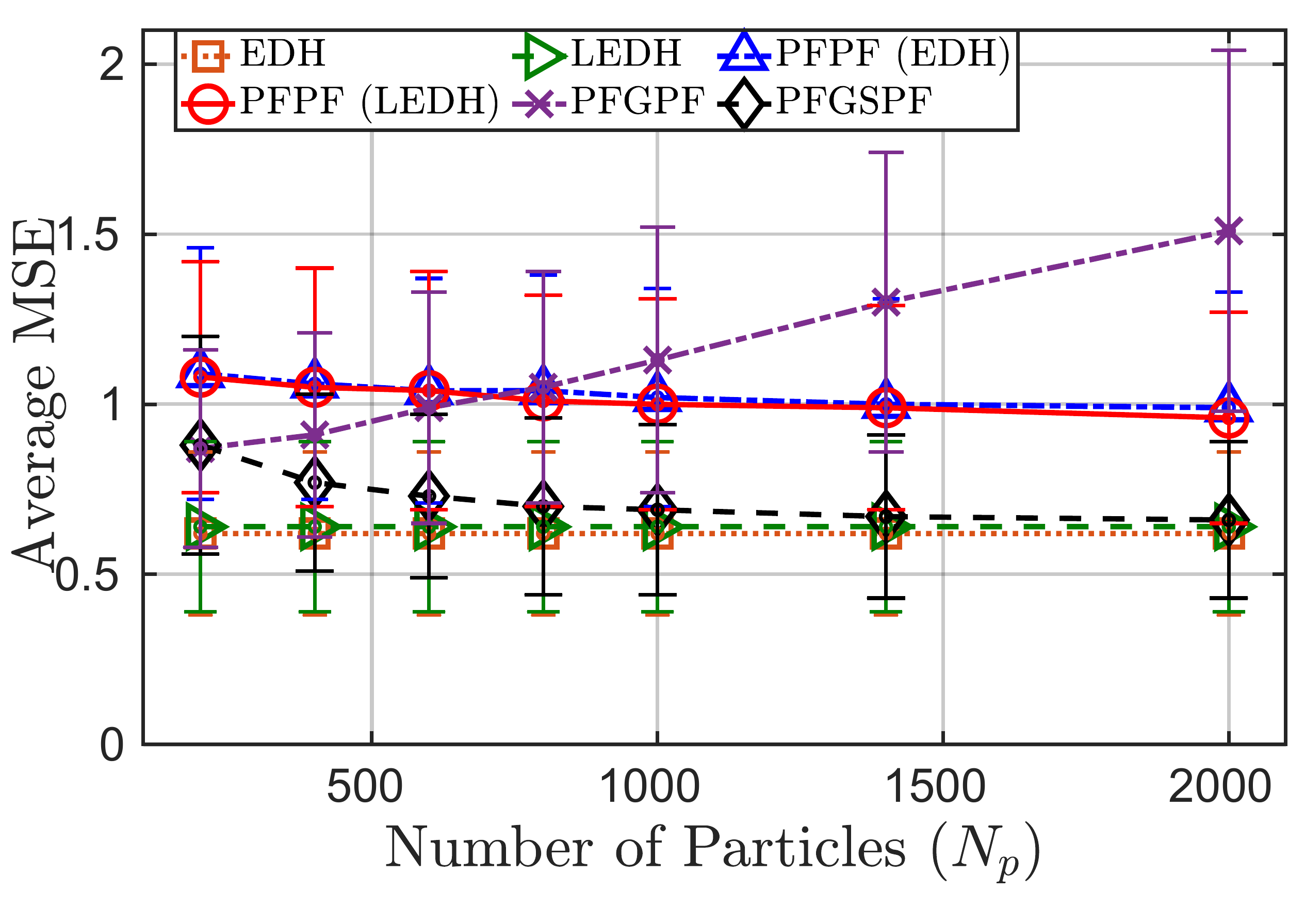}
\caption{Average MSE vs $N_p$ in the large spatial sensor networks simulation. For PFGSPF, $N^{*}_p = 200$ is fixed and $G$ is varied. Error bars indicate standard deviation.}
\label{fig:Septier_avg_MSE}
\end{figure}

We also examine the filters' performance in $144$-dimensional large spatial sensor networks used in \cite{septier2015langevin,PFGPF2022}. The dynamic model follows Generalized Hyperbolic (GH) skewed-t distribution, a heavy-tailed distribution \cite{zhu2010generalized}, where the model equations and parameter values used in the simulation can be found in \cite{PFGPF2022}. Each experiment is simulated for 30 time steps and we conduct the simulations for 100 times.

Figure \ref{fig:Septier_avg_MSE} shows the average MSE of all the filters for different $N_p$ over 100 simulations. We observe that the EDH has the lowest average MSE and standard deviation compared to other filters. As we increase the number of Gaussians $G$, the performance of the PFGSPF improves and approaches that of the EDH.

The PFGSPF is implemented with different values of $N^{*}_{p}$ and $G$ as shown in Table \ref{tab:septier_144_avg_MSE_all_filters}. We observe that, for fixed $N^{*}_{p}$, an increase in the number of Gaussians $G$ improves its performance. Meanwhile, the filter performance degrades as we increase $N^{*}_p$ while keeping $G$ constant. The same behaviour is observed for the PFGPF in Figure \ref{fig:Septier_avg_MSE}. We speculate that as $N^{*}_{p}$ increases, more particles may fall into the regions of posterior with less probability, resulting in poorer approximation.

\begin{table}[t]
\small
\centering
    \begin{tabular}{|c|c|c|c|c|c|c|c|c|c|}
    \hline
    {$N_p$} & 
      \multicolumn{2}{c|}{Average MSE [avg $\pm$ sd]}
       \\
    \cline{2-3}
    & PFGSPF ($N^{*}_{p} \times G$) & PFGSPF ($N^{*}_{p} \times G$) \\
    \hline
     200& 0.88 $\pm$ 0.32 (200x1) &- \\
    \hline
     400& 0.77 $\pm$ 0.26 (200x2)& 0.77 $\pm$ 0.26 (200x2)   \\
    \hline
     600&0.73 $\pm$ 0.24 (200x3)& 0.79 $\pm$ 0.28 (300x2)  \\
    \hline
     800&0.70 $\pm$ 0.26 (200x4)& 0.81 $\pm$ 0.29 (400x2)  \\
    \hline
     1000& 0.69 $\pm$ 0.25 (200x5)& 0.84 $\pm$ 0.29 (500x2)    \\
    \hline
     1400& 0.67 $\pm$ 0.24 (200x7)& 0.73 $\pm$ 0.27 (350x4)    \\
    \hline
      2000& 0.66 $\pm$ 0.23 (200x10)& 0.77 $\pm$ 0.28 (500x4)   \\
    \hline
    \end{tabular}
 \caption{\label{tab:septier_144_avg_MSE_all_filters} Average MSE in the large spatial sensor networks simulation.}
  \end{table}

\begin{figure}[h]
\centering
\includegraphics[width=0.45\textwidth]{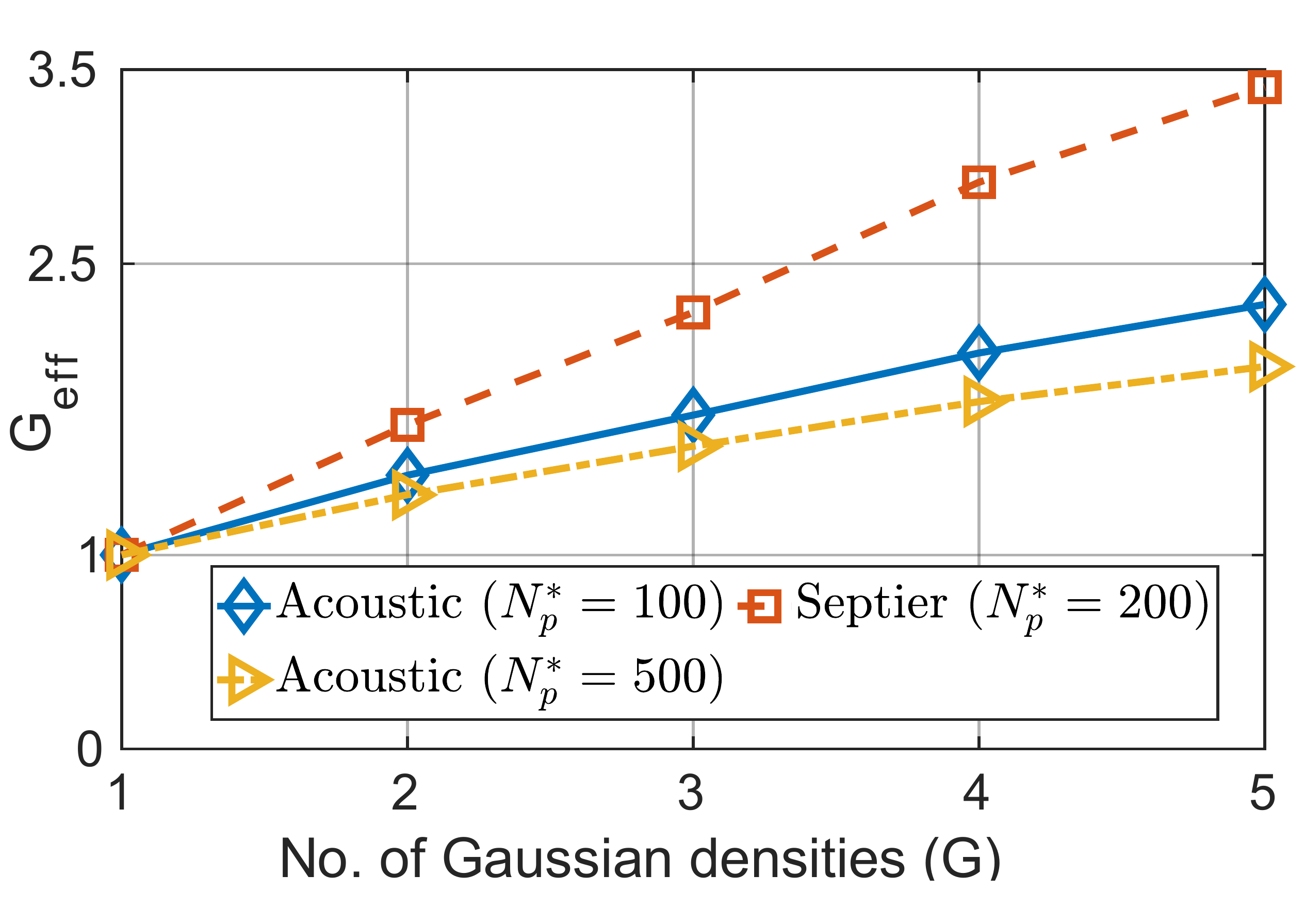}
\caption{\label{tab:effective_Gaussains} Average $G_\text{eff}$ of PFGSPF. (Acoustic: 500 simulations, large spatial sensor network: 100 simulations).}
\label{fig:tracks}
\end{figure}

Figure \ref{tab:effective_Gaussains} shows the average effective number of Gaussians in the two numerical simulations for different values of $G$ in the PFGSPF. In acoustic simulation, the average effective number of Gaussians is  $2$ when $G = 5$ are available. In the large spatial sensor network simulation, the average effective number of Gaussians is $3.5$ when $G = 5$ are available.

  
  
\vspace{-0.6cm}
\section{Conclusion}
\label{sec:conclusion}
\vspace{-0.1cm}
In this paper, we proposed the particle flow Gaussian sum particle filter (PFGSPF) constructed using a bank of particle flow Gaussian particle filters. It leverages particle flow to construct more effective proposals and Gaussian sum particle filters to approximate multi-modal high-dimensional distributions. We apply the filter in complex high-dimensional numerical simulations to demonstrate the effectiveness of the proposed method. In particular, we observe that, for fixed number of particles per Gaussian, an increase in the number of Gaussian components results in improved performance.

\vfill\pagebreak

\small
\bibliographystyle{IEEEbib}
\bibliography{references.bib}

\end{document}